\documentclass[twocolumn,superscriptaddress,prl,amsmath,amssymb]{revtex4-2}

\usepackage{graphicx}
\usepackage{color}
\usepackage{amsmath}
\usepackage{bm}
\usepackage[normalem]{ulem}
\usepackage{braket}
\usepackage[
  colorlinks=true,
  linkcolor=blue,
  citecolor=blue,
  urlcolor=blue
]{hyperref}
\DeclareMathOperator{\Tr}{Tr}

\begin{document}

\title{Quantum Geometric Limits for Non-Abelian Holonomies}

\author{François Impens}
\affiliation{Instituto de F\'{\i}sica, Universidade Federal 
Rio de Janeiro, 21941-972 Rio de Janeiro, RJ, Brazil}

\author{David Guéry-Odelin}
\affiliation{Laboratoire Collisions Agrégats Réactivité, UMR 5589, FERMI, Université de Toulouse, CNRS, 118 Route de Narbonne, 31062 Toulouse CEDEX 09, France.}

\date{\today}


\begin{abstract}
Stokes' theorem turns Abelian Berry phases into curvature fluxes, whereas
path ordering precludes such a simple formula for non-Abelian holonomies.
We show that a quantitative form of this intuition survives: arbitrary
Wilczek--Zee holonomies obey a universal quantum geometric limit~(QGL), in
which the holonomy magnitude is bounded by a surface integral of the
non-Abelian curvature norm. Recasting holonomic evolution as an effective
Stokes--Schr\"odinger dynamics driven by transported curvature, we identify
the QGL as the geometric counterpart of conventional quantum speed limits,
with a time-integrated generator norm replaced by a surface-integrated
curvature cost. The induced contour--surface variational problem is governed
by a non-Abelian Lorentz force, which we address with a brachistochrone ansatz
of curvature-weighted geodesics. Applied to an SU(2) tripod dark subspace,
near-optimal protocols spontaneously align the transported curvature along a
single Lie-algebra direction, effectively taming non-Abelianity.
%
\end{abstract}


\maketitle

Geometric phases reveal that quantum evolution is not governed solely by
dynamical phases, but also by the geometry of the path followed in parameter
space \cite{Pancharatnam1956,simon1983,berry1984,anandan1988}. In the
adiabatic regime, this geometric structure takes the form of parallel
transport generated by a gauge connection and its associated curvature
\cite{nakahara2003,bohm2003}. For a nondegenerate state the resulting
holonomy reduces to the Abelian Berry phase. In the presence of degeneracies,
however, the transported eigenspace is multidimensional, and the adiabatic
connection becomes matrix-valued. Closed paths then generate non-Abelian
Wilczek--Zee holonomies, which act as unitary transformations within the
degenerate subspace and provide the basis of holonomic quantum gates
\cite{wilczek1984,Uhlmann1986,zanardi1999,nakahara2003,chruscinski2004,Dai25}. 


This geometric viewpoint has become increasingly relevant experimentally: non-Abelian geometric phases and holonomic transformations have been explored on a broad range of quantum platforms, including   trapped ions~\cite{duan2001,leibfried2003}, cold atoms~\cite{Ruseckas2005,Leroux18,Bharath2019,Sugawa2021,WilkowskiPRL22,Madasu2025}, superconducting circuits~\cite{abdumalikov2013,xu20}, solid-state spins~\cite{zu2014,arroyo-camejo2014,yale2016} and photonic systems~\cite{IadecolaSchusterChamon2016,Sun2022NonAbelianThouless,Parto2023,Chen2025HighDimensionalHolonomy,Guillot25}.
 These developments naturally raise the question of the fundamental geometric
resources required to implement a prescribed holonomy. The Abelian Berry phase, equal to the curvature flux through any surface bounded by the loop, provides the basic geometric intuition. In the
non-Abelian case, however, the situation is fundamentally subtler. Path
ordering prevents the logarithm of the holonomy from being written as a
simple curvature flux. The central question is therefore whether one can nevertheless derive a universal geometric bound for non-Abelian holonomies, and formulate the corresponding optimization as a variational holonomic control problem.


This is the program of the present work. In ordinary quantum speed limits~(QSLs), the distance travelled in Hilbert or unitary space is bounded by the time integral of a dynamical generator norm, typically characterized by the energy uncertainty~\cite{Mandelstam1945,QSLAnandan90,Margolus1998,Maccone03,Taddei2013QSL,DelCampo2013OpenQSL,
Deffner2013NonMarkovianQSL,deffner2017quantum,Shanahan2018QSLClassicalTransition,QSLImpens21,
Ness2021QSLExperiment,QSLImpens25,Nelson26}.
Recently, isoholonomic QSLs have been derived for cyclic quantum evolutions, bounding the length of one-dimensional trajectories in state or density-operator space~\cite{Sonnerborn26}. Unlike these QSLs, holonomic transformations generated by Wilczek--Zee connections arise from closed loops in control-parameter space. This motivates the concept of a quantum geometric limit~(QGL), in which the dynamical resource of conventional QSLs is replaced by a geometric surface density determined by the non-Abelian curvature and integrated over a spanning surface bounded by the loop. In practice, a minimal time scale would emerge in the presence of constraints on control velocities, spectral gaps, or admissible adiabatic errors.


We begin by deriving a differential form of the
non-Abelian Stokes theorem~\cite{HalpernStokes1,Stokes2,KondoStokes3} adapted to holonomic evolution. By encoding a line integral into the geometric generator, this representation restores an effective one-dimensional
evolution equation naturally leading to universal geometric bounds for
holonomies. For a prescribed target
holonomy \(U^*\), the resulting QGL defines a geometric cost for any spanning surface and turns
the search for optimal implementations into a coupled variational
problem over contours \(\mathcal C\) and spanning surfaces
\(\Sigma(\mathcal C)\) bounded by  \(\mathcal C\). Beyond the bound itself, this variational formulation provides a constructive
route toward protocols approaching the geometric limit. While the full
problem is nonlocal and very challenging, we adapt the time-optimal framework
of quantum brachistochrones~\cite{Carlini06,Carlini07,Mostafazadeh07} to
construct approximate near-optimal solutions in the non-Abelian holonomic
setting. The resulting candidate contours satisfy driven geodesic equations in
a curvature-weighted metric, with a Lorentz-force term generated by the
non-Abelian curvature. We illustrate these principles in the degenerate dark subspace of a
four-level tripod atomic system.  More broadly, the framework applies
to adiabatically controlled systems  supporting degenerate subspaces,
including \(\Lambda\) and tripod configurations encountered in quantum
optics and atomic physics~\cite{fleischhauer2005,unanyan99,Molmer07,sjoqvist2012}.


%

\noindent \textit{Non-Abelian transport and Fubini--Study metric.} We consider the non-Abelian adiabatic transport of a quantum state $| \psi(\boldsymbol{\lambda}(t)) \rangle = \sum_{n=1}^N c_n(\boldsymbol{\lambda}(t)) | D_n (\boldsymbol{\lambda}(t)) \rangle$
within an $N$-dimensional degenerate subspace $\mathcal{H}_N$ along a closed path $\mathcal{C}$ in parameter space.
This evolution generates a unitary gate
\begin{equation}
\label{eq:paralleltranport}
U(\mathcal C)=\mathcal P\exp\!\left(i\oint_{\mathcal C} A_\mu\,d\lambda^\mu\right)
\end{equation}
acting on the initial coefficient vector $ \boldsymbol{c}(\boldsymbol{\lambda}(0))$. $A_\mu(\boldsymbol{\lambda})_{ab}
=
i\,\langle D_a(\boldsymbol{\lambda}) \mid \partial_\mu D_b(\boldsymbol{\lambda}) \rangle$ is the Wilczek--Zee connection.
The path-ordering $\mathcal P$ is essential as generally
$[A_\mu(\boldsymbol{\lambda}),A_\nu(\boldsymbol{\lambda}')] \neq 0$,
and $U(\mathcal C)$ depends on the full path $\mathcal C$. The choice of basis $\{ | D_n (\boldsymbol{\lambda}) \rangle \}$ along the path defines an $SU(N)$ gauge freedom. For closed paths, \(U(\mathcal C)\) transforms only by a single conjugation under local basis changes. Its conjugacy class then encodes gauge-invariant properties
of the non-Abelian holonomy, in analogy with the Berry phase. The variation of the degenerate subspace with the control parameters induces
a natural Riemannian metric on the parameter manifold \(\mathcal M\). Let $P(\boldsymbol{\lambda})=\sum_{a=1}^{N}|D_a(\boldsymbol{\lambda})\rangle\langle D_a(\boldsymbol{\lambda})|$ denote the projector onto $\mathcal{H}_N(\boldsymbol{\lambda})$. The Fubini--Study metric on the manifold of degenerate subspaces is then defined as
$ds^2 = \tfrac12 \mathrm{Tr}(dP\,dP)=g_{\mu\nu}(\boldsymbol{\lambda})\,d\lambda^\mu d\lambda^\nu$,
which is gauge-invariant and measures the distinguishability of nearby subspaces~\cite{Carlini07}. The metric $g_{\mu\nu}$ provides the geometric structure underlying both the QGL framework developed below and the quantum brachistochrone formulation.

 \noindent\textit{Curvature flux and Berry phase.}
In the Abelian case, the geometric cost of a closed adiabatic cycle is directly controlled by curvature flux. Indeed, Stokes' theorem gives \(\Phi[\mathcal C]=\iint_{\Sigma(\mathcal C)}F_{\mu\nu}(\boldsymbol\lambda)d\Sigma^{\mu\nu}\), for any surface \(\Sigma(\mathcal C)\) bounded by \(\mathcal C\). If \(|F_{\mu\nu}|\leq F_{\rm max}\) on the accessible parameter manifold \(\mathcal S\), one obtains \(\mathcal A_\Sigma\geq |\Phi[\mathcal C]|/F_{\rm max}\). For a spin-\(\tfrac12\) in a slowly varying magnetic field, \(H=-\boldsymbol B\cdot\boldsymbol\sigma\), with \(\boldsymbol B=B(\sin\theta\cos\phi,\sin\theta\sin\phi,\cos\theta)\), the curvature satisfies \(|F_{\theta\phi}|\leq 1/2\). Thus a Berry phase \(\Phi[\mathcal C]=\pi\) requires \(\mathcal A_\Sigma\geq 2\pi\), i.e. at least a hemisphere of the Bloch sphere, and this bound is saturated because \(\Phi[\mathcal C]=\Omega(\Sigma)/2\). This Abelian example illustrates the emergence of a minimal geometric cost associated with holonomic evolution.

\noindent \textit{Effective Stokes-Schr\"odinger dynamics.}
We construct a non-Abelian Stokes representation~\cite{HalpernStokes1,DiakonovStokes2,KondoStokes3} specifically adapted for the present geometric setting. More precisely, the holonomy problem can be recast as an effective one-dimensional
Schr\"odinger equation in the Stokes representation~\cite{SM}
\begin{equation}
\label{eq:stripdynamics}
\partial_{s_1} V = - i \mathcal{K}(s_1) V
\end{equation}
where $V(s_1)$ is a unitary evolution operator satisfying
$V(0)=\mathbb{I}$ and $V(1)=U(\mathcal{C})$.
The strip generator $\mathcal K(s_1) $ acts as an effective Hamiltonian and is constructed from the 
non-Abelian curvature transported to a common base point on the surface~\cite{SM}. Explicitly,  $\mathcal K(s_1) =
\int_0^1 ds_2\;\widetilde{F}_{s_1s_2}(s_1,s_2)
$ in terms of the projected curvature $\widetilde{F}_{s_1s_2}(s_1,s_2)=\widetilde{F}_{\mu \nu}(\boldsymbol{\lambda}_\Sigma(s_1,s_2)) \partial_{s_1} \lambda_{\Sigma}^{\mu} \partial_{s_2} \lambda_\Sigma^{\nu}$. The latter captures the geometry of the parametrized surface $\boldsymbol{\lambda}_{\Sigma}(s_1,s_2)$ through the tangent vectors $\partial_{s_1} \boldsymbol{\lambda}_\Sigma$ and $\partial_{s_2} \boldsymbol{\lambda}_\Sigma$. 
Equation~\eqref{eq:stripdynamics} yields a natural QSL-like framework, in which the relevant resource is the curvature encoded in $\mathcal{K}(s_1)$ instead of the energy variance associated with a standard Hamiltonian. {The resulting QGL originates from the geodesic distance associated with
the effective unitary evolution~\eqref{eq:stripdynamics}, in close
analogy with ordinary QSLs, which arise from geodesic distances in the
Fubini--Study geometry~\cite{QSLAnandan90}.} Unlike ordinary Hamiltonian dynamics, however, the strip generator
$\mathcal K(s_1)$ itself is obtained from a line integral -- thus leading to QGLs given by surface integrals.



\noindent\textit{Hilbert--Schmidt QGL for non-Abelian holonomies.}
Starting from the Stokes--Schr\"odinger equation~\eqref{eq:stripdynamics}, one
can first transpose the QSL derivation by Vaidman~\cite{Vaidman92}, recently extended to unitary gates~\cite{QSLImpens25},
to the QGL setting. This approach, based on elementary arguments, yields a Mandelstam--Tamm QGL relating
the integrated curvature norm to the overlap of the target holonomy with the identity~\cite{SM}:
\begin{equation}
\label{eq:MTbound}
\arccos\left|
\frac1N{\rm Tr}\,U(\mathcal C)
\right|
\le
\iint_{\Sigma(\mathcal C)} \|F(\boldsymbol{\lambda})\|_{\rm HS} \: dS.
\end{equation}
Here the overlap reduces to a simple trace. From now on, we use the Hilbert--Schmidt norm
$\|A\|_{\rm HS}=\sqrt{\langle A,A\rangle}$ with inner product $\langle U,V\rangle=\mathrm{Tr}(U^\dagger V)/N$. Also, $\|F(\boldsymbol{\lambda})\|_{\rm HS}
=
\sqrt{\frac12\langle F_{\mu\nu},F^{\mu\nu}\rangle }$. In the QGLs~(\ref{eq:MTbound}--\ref{eq:tightbound_Abelian}), \(dS\)
is the gauge-invariant area element induced on \(\Sigma(\mathcal C)\)
by the Fubini--Study metric~(see \cite{SM}).

Considering instead the geodesic distance leads to a more fundamental bound for any loop $\mathcal C$ and any smooth spanning surface $\Sigma(\mathcal C)$~(see \cite{SM}):
\begin{equation}
\|\log_{\min} U(\mathcal C)\|_{\rm HS}
\le
\iint_{\Sigma(\mathcal C)}
\|F(\boldsymbol{\lambda})\|_{\rm HS}\,dS .
\label{eq:nonabelian_area_bound_main}
\end{equation}
$\log_{\min}U$ denotes the minimal logarithm, and $\Theta(U)\equiv\|\log_{\min}U\|_{\rm HS}$ defines a gauge-invariant measure of gate magnitude, reducing to the usual rotation angle for $SU(2)$. Equation~(\ref{eq:nonabelian_area_bound_main}) defines an intrinsic  QGL and sets a minimal geometric cost for generating a prescribed holonomy. The associated resource is the non-Abelian curvature-norm density integrated over a spanning surface. For a fixed holonomy $U(\mathcal C)$, the bound~\eqref{eq:nonabelian_area_bound_main} holds for any smooth spanning surface $\Sigma(\mathcal C)$ and can therefore be optimized over the surface choice. More generally, the minimal geometric cost of a target gate $U^*$ is obtained by minimizing the right-hand side over all loops $\mathcal C$ satisfying $U(\mathcal C)=U^*$ and all spanning surfaces $\Sigma(\mathcal C)$. This defines a coupled optimization problem over contours and surfaces, which admits a variational formulation discussed below.


\noindent \textit{Operator-norm QGL for non-Abelian holonomies.}
 A tighter universal QGL bound follows from Eq.\eqref{eq:stripdynamics}~\cite{SM},
\begin{equation}
\label{eq:tightbound_Abelian}
\|\log_{\min} U(\mathcal C)\|_{\rm HS}
 \leq \iint_{\Sigma(\mathcal C)} \| F(\boldsymbol{\lambda})\|_{\rm op}\, dS .
\end{equation}
This bound is obtained by viewing the non-Abelian curvature as a map acting on antisymmetric bivectors, $T_F:B^{\mu \nu} \mapsto F_{\mu \nu}(\boldsymbol{\lambda}) B^{\mu \nu}/2$. Precisely, $\| F(\boldsymbol{\lambda})\|_{\rm op}$ denotes the operator norm $\|F\|_{\rm op}:=\sup_{B\neq0} \|\tfrac12 F_{\mu\nu}B^{\mu\nu}\|_{\rm HS}/ \|B\|_{\rm vec}$,
with $\|B\|_{\rm vec}=\sqrt{ \tfrac{1}{2} B_{\mu \nu} B^{\mu \nu}}$. At the price of involving a less tractable geometric
norm, the operator-norm bound~\eqref{eq:tightbound_Abelian} is always as tight as, or tighter than, the
Hilbert--Schmidt bound~\eqref{eq:nonabelian_area_bound_main}
~\cite{SM}. It therefore defines the sharpest QGL for non-Abelian
holonomies, while the relaxed Hilbert--Schmidt
QGL~\eqref{eq:nonabelian_area_bound_main} remains particularly useful
for the construction of near-optimal protocols.

\noindent\textit{Surfaces with effectively Abelian transported curvature.} We now consider the special case of spanning surfaces
$\Sigma(\mathcal C)$ for which the transported curvature field remains
everywhere collinear with a fixed Lie-algebra generator. The existence of
such surfaces depends both on the underlying curvature field and on the
chosen contour. In this
regime, the strip generators $\mathcal K(s_1)$ commute, path ordering
drops out from Eq.~\eqref{eq:stripdynamics}, and the holonomy reduces to
an ordinary exponential. For such surfaces, the obstruction associated with non-commutativity is
lifted, and the bound reduces to a direct non-Abelian generalization
of the Berry-curvature flux relation~\cite{SM}, with the
Hilbert--Schmidt norm replacing the absolute value.

\noindent\textit{Non-Abelianity and QGL saturation.}
In the Abelian case, saturation of the geometric bound is achieved by
choosing a surface whose local orientation is everywhere adapted to the
curvature field. For generic non-Abelian holonomies, the situation is more
subtle, since variations in the Lie-algebra direction of the transported
curvature hinder a globally coherent flux accumulation. Nevertheless, in
the examples considered below, near-optimal solutions exhibit an 
approximate alignment of the transported curvature over the optimized
surface $\Sigma(\mathcal C)$.

\noindent \textit{Non-Abelian QGL optimization.}  
We formulate the QGL optimization as a variational problem minimizing the Hilbert--Schmidt bound~\eqref{eq:nonabelian_area_bound_main} for a target gate $U^*$. We introduce the action
$
S[\boldsymbol{\lambda}_{\Sigma},\boldsymbol{\lambda}_{\mathcal{C}},K_0]
= S_{\rm Geo}[\boldsymbol{\lambda}_{\Sigma}] + S_{U^*}[K_0,\boldsymbol{\lambda}_{\mathcal{C}}],
$
where $S_{\rm Geo}[\boldsymbol{\lambda}_{\Sigma}] = \iint_{\Sigma(\mathcal{C})} \|F\|_{\rm HS} dS$ measures the geometric cost, and $S_{U^*}[K_0,\boldsymbol{\lambda}_{\mathcal{C}}]$ enforces the target holonomy.
The contour $\boldsymbol{\lambda}_{\mathcal C}$ determines the holonomy, while the surface $\boldsymbol{\lambda}_{\Sigma}$ controls the geometric cost, leading to a coupled optimization over paths and surfaces. Here $K_0\in \mathfrak{su}(N)$ is a constant anti-Hermitian Lie-algebra multiplier acting as a geometric generator. 

Stationarity with respect to $K_0$ enforces $U(\mathcal{C})=U^*$, while variation with respect to the surface and contour yields bulk and boundary conditions (see~\cite{SM}). We henceforth parametrize the contour by $t\in[0,T]$, with $s_1\equiv t/T$. Variation yields the boundary equation for the optimal contour:
\begin{equation}
\|F(\boldsymbol{\lambda}_{\mathcal C}(t))\|_{\rm HS} \,\hat{t}_\mu
= 
\mathrm{Re}\,\Tr\!\Big[
i\,K(t)\,
F_{\mu\nu}(\boldsymbol{\lambda}_{\mathcal C}(t))\,
\dot\lambda_{\mathcal{C}}^\nu(t)
\Big],
\label{eq:equation_of_motion_minimal_flux}
\end{equation}
where $\hat{t}^\mu$ is the outward surface traction vector and $K(t)=U(t)^{-1}K_0 U(t)$ is the adjoint-transported generator along the contour.
Equation \eqref{eq:equation_of_motion_minimal_flux} has the form of a non-Abelian Lorentz-force law for the boundary loop: the left-hand side describes an effective surface tension, while the right-hand side corresponds to a gauge-field force induced by the holonomy.

\noindent\textit{Effective QGL optimization via non-Abelian quantum brachistochrone} -- The coupled loop--surface equations define a challenging nonlocal problem.
As an effective approximation, we use the geometric intuition that shorter
boundary loops provide natural candidates for lower
spanning-surface costs. To capture, at the contour level, the curvature-norm weighting of the QGL
cost, we introduce the conformally rescaled metric
\(g'_{\mu\nu}=\|F\|_{\rm HS}g_{\mu\nu}\). We thus solve the corresponding
brachistochrone problem under the holonomy constraint \(U(T)=U^*\) and a
closure constraint, and then minimize the geometric functional
\(S_{\rm Geo}[\boldsymbol{\lambda}_\Sigma]\) over all spanning surfaces bounded by the
resulting loop. This provides an effective decoupling of the
complete loop--surface problem.

To obtain contour candidates, we formulate a non-Abelian
brachistochrone problem in the curvature-weighted metric. In the standard
brachistochrone formulation, among all paths \(\boldsymbol{\lambda}(t)\)
implementing the target holonomy \(U^*\), one seeks the path of shortest
Fubini--Study length $\mathcal{L}= \int_0^T  dt\sqrt{g_{\mu\nu}(\boldsymbol{\lambda})\,\dot\lambda^\mu\,\dot\lambda^\nu} $. For a fixed time interval, this is equivalent to minimizing the energy functional
\begin{equation}
J_L[\boldsymbol{\lambda},\dot{\boldsymbol{\lambda}}]
=
\frac12
\int_0^T \!dt\;
g_{\mu\nu}(\boldsymbol{\lambda})\,\dot\lambda^\mu\,\dot\lambda^\nu,
\label{eq:cost_energy}
\end{equation}
under the constraint $U(T)=U^*$~\cite{Ansel2024QOC}, with the path velocities $\dot{\boldsymbol{\lambda}}$ as controls. The minimizing parametrization has constant Fubini–Study speed, yielding $J_L[\boldsymbol{\lambda},\dot{\boldsymbol{\lambda}}]=\mathcal{L}^2/(2T)$. For the QGL contour ansatz, the same construction is applied in the
curvature-weighted metric \(g'_{\mu\nu}\). Applying the Pontryagin maximum principle~\cite{pontryagin1962} yields a driven geodesic equation~(see~\cite{SM}),
\begin{equation}
\ddot{\lambda}^\mu + \Gamma^\mu_{\alpha\beta}[g'] \dot{\lambda}^\alpha \dot{\lambda}^\beta 
=  \mathcal{D}_{\rm QGL}^{\mu \nu (K)}(\boldsymbol{\lambda},t)\, \dot{\lambda}_\nu.
\label{eq:contour_equation}
\end{equation}
$\mathcal{D}_{\rm QGL}^{\mu \nu (K)} = i \langle K(t), F^{\mu\nu} \rangle / \|F\|_{\rm HS}$ denotes the projection of the normalized curvature onto the transported generator $K(t)$. Since $K(t)$ is anti-Hermitian, the factor $i$ ensures that $\mathcal{D}_{\rm QGL}^{\mu \nu (K)}$ is real. This equation has the structure of a Lorentz-force law on the control manifold endowed with the metric $g'_{\mu\nu}$, with $\mathcal{D}_{\rm QGL}^{\mu \nu (K)} \dot{\lambda}_\nu$ acting as an effective force.
The curvature sets the local direction of the field, while $K(t)$ selects the corresponding Lie-algebra component and plays the role of an effective internal charge. Finally, when \(\|F(\boldsymbol{\lambda})\|_{\rm HS}\) is constant, the
conformal factor reduces to a global rescaling, and the loop candidates for QGL optimization coincide with the non-Abelian brachistochrones in the Fubini--Study metric.

\noindent\textit{Application to the quantum tripod.} 
We illustrate the framework in a paradigmatic non-Abelian system~\cite{Ruseckas2005,Leroux18,SM}: a tripod configuration with three degenerate ground states coupled to a single excited state. The dynamics is restricted to a two-dimensional dark subspace $\mathcal D(t)$ spanned by $|D_j\rangle=e^{i\chi/2}|\tilde D_j\rangle$, with
$| \tilde{D}_1 \rangle =  \sin\phi\,e^{-\frac {i} {2} \varphi}\ket{1}-\cos\phi\,e^{\frac {i} {2} \varphi}\ket{2}$ and 
$| \tilde{D}_2 \rangle =    \cos\theta\cos\phi\,e^{-\frac {i} {2} \varphi}\ket{1}
+\cos\theta\sin\phi\,e^{\frac {i} {2} \varphi}\ket{2}-\sin\theta\,\ket{3}$.
The control parameters are $\boldsymbol{\lambda}(t)=(\theta,\phi,\varphi)$, while $\chi$ generates a trivial gauge transformation.  The connection reads $A_\theta = 0_{2\times 2}$, $A_\phi= \cos\theta\,\sigma_y,$ $A_{\varphi}
=
\frac{1}{2}\sin (2\phi)\cos\theta \, \sigma_x
\;-\;
\frac{1}{4}\cos(2\phi)\left(1+\cos^{2}\theta\right)\sigma_z$, yielding a genuinely non-Abelian structure $[A_\phi,A_\varphi]\neq 0$.

\begin{figure}[t]
    \centering
    \includegraphics[width=7 cm]{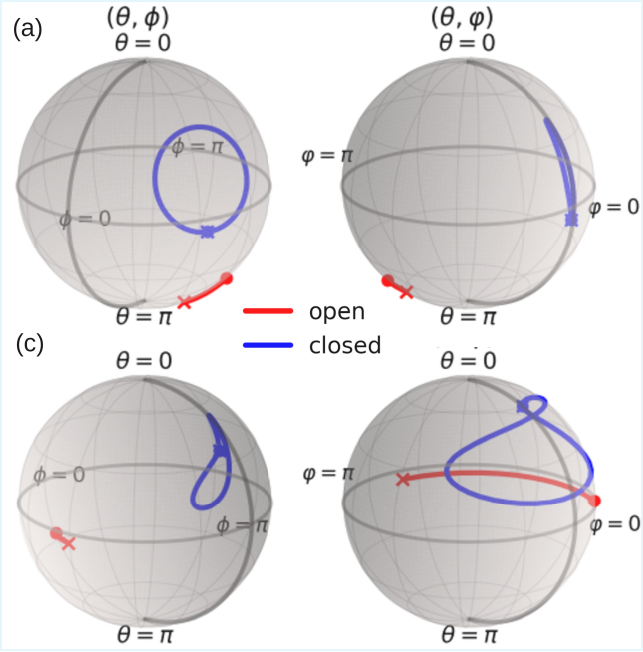}
    \caption{Non-Abelian brachistochrones:  Open~(red) and closed~(blue) brachistochrones for (a,b) $U_1^*=e^{i \frac {\pi} {3} \sigma_y}$ and  (c,d) $U_2^*=e^{i \frac {\pi} {3} \sigma_z}$. (a,c) show the $(\theta,\phi)$ motion on a sphere, while (b,d) represent the $(\theta,\varphi)$ motion. Open brachistochrones follow trajectories close to latitude circles. Target holonomies are reached within $\epsilon=1-\mathcal{F}<10^{-5}.$ ($\mathcal{F}= |\langle U^*, U(T) \rangle|^2$) }
    \label{fig:Brachistochrone}
\end{figure}

As \(\|F\|_{\rm HS}=\sqrt{3}\) everywhere in the tripod model, we obtain closed
candidate loops by solving the quantum brachistochrone in the original
Fubini--Study metric \(g_{\mu\nu}\). We also compute the corresponding open
brachistochrone paths for comparison. We use a shooting method on the parameter set $\{ \boldsymbol{\lambda}_0, \dot{\boldsymbol{\lambda}}_0, K_0,T \}$  that enforces both gate fidelity and closure of the parameter trajectory. Figure~\ref{fig:Brachistochrone} shows the resulting trajectories for two target gates, $U_1^*=e^{i \tfrac{\pi}{3} \sigma_y}$ and $U_2^*=e^{i\tfrac{\pi}{3}\sigma_z}$. For $U_1^*$, the optimal open trajectory moves predominantly along $\phi$, consistent with $A_\phi\propto\sigma_y$, while keeping $\varphi$ nearly constant to suppress unwanted $\sigma_x$ and $\sigma_z$ rotations. For $U_2^*$, the optimal open trajectory follows the equator $\theta=\pi/2$, where $A_\varphi\propto\sigma_z$. Since $A_\theta=0$, one may expect optimal trajectories to move at fixed $\theta$, which is approximately realized for open trajectories. Closed holonomies, however, require a finite enclosed area and therefore nonzero curvature flux, forbidding motion along a single coordinate. The associated paths explore at least two directions: $(\theta,\phi)$ for $U_1^*$ and $(\theta,\varphi)$ for $U_2^*$.
The corresponding Fubini--Study lengths are
\(\mathcal L_{1,2}^{\rm(open)}\simeq0.4\) for open-path implementations,
whereas the closed-loop trajectories are substantially longer,
\(\mathcal L_1^{\rm(closed)}=2.5\) and
\(\mathcal L_2^{\rm(closed)}\simeq2.7\). This highlights the geometric overhead of closed holonomic implementations relative to open-path non-Abelian gates.

 To probe the impact of non-Abelianity on geometric efficiency, we next investigate the tightness of the bounds~\eqref{eq:nonabelian_area_bound_main} and~\eqref{eq:tightbound_Abelian}. We consider the family of target gates $U^*(\alpha)= e^{i \Phi \hat{\boldsymbol{n}}(\alpha) \cdot \boldsymbol{\sigma}},$ with fixed angle $\Theta(U^*)= \Phi =\pi/4$ and $\hat{\boldsymbol n}(\alpha)=\cos\alpha\,\hat{\mathbf y}
+\sin\alpha\,\hat{\mathbf z}$. For each target gate, we first construct a closed contour $\mathcal C$ satisfying the holonomy constraint and then determine an associated spanning surface $\Sigma(\mathcal C)$ minimizing the curvature flux through direct relaxation of the flux functional.

\begin{figure}[t]
    \centering
    \includegraphics[width=9 cm]{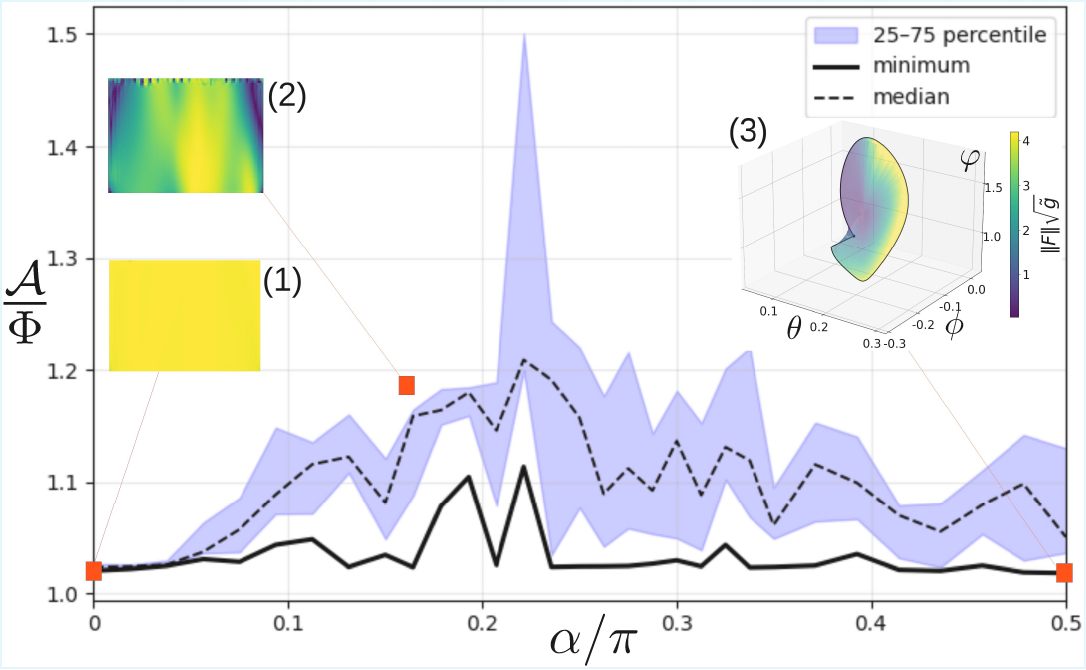}
    \caption{QGL efficiency $\eta=\mathcal{A}(\Sigma(\mathcal{C}))/\Phi$ as a function of the rotation-axis orientation for the target gate
$U^*(\alpha)=e^{i\Phi\,\hat{\boldsymbol n}(\alpha)\cdot\boldsymbol{\sigma}}$,
with fixed angle $\Phi=\pi/4$ and
$\hat{\boldsymbol n}(\alpha)=\cos\alpha\,\hat{\mathbf y}+\sin\alpha\,\hat{\mathbf z}$.
For each axis orientation, the optimization was repeated 10 times from different initial seeds. The solid black line shows the lowest-cost solution found for each angle, the dashed line the median over all runs, and the blue shaded region the dispersion across runs. Insets (1,2) display the scalar field
$\hat{\mathbf v}_{\sigma}(s_1,s_2)\!\cdot\!\hat{\mathbf v}_0$,
which measures the local alignment of the transported curvature over the optimal surface. Here $\hat{\mathbf v}_{\sigma}(s_1,s_2)$ is the unit vector associated with the Lie-algebra direction of the transported curvature projected onto the local surface element, while $\hat{\mathbf v}_0$ denotes its dominant orientation over the optimal surface. A nearly uniform inset indicates that the transported curvature is almost collinear with a single generator, corresponding to near-saturation of the Abelian bound. The 3D inset~(3) shows a representative optimal contour and surface for the gate $U_2^*=U^*(\pi/2)$ (angles in units of $\pi$), where $\|F\|\equiv\|F\|_{\mathrm{HS}}$ and $\sqrt{\tilde g}$ determines the local area element.}
\label{fig:QSLOptimization}
\end{figure}

 While still more efficient solutions may exist, the present scheme provides explicit constructive realizations and corresponding upper bounds on the minimal geometric cost. Our numerical results suggest an optimization landscape with multiple competing branches satisfying the same closure and holonomy constraints. Since $\Theta(U^*)= \Phi$, the efficiency ratio $\eta=\mathcal A(\Sigma(\mathcal C))/\Phi$ provides a direct measure of the tightness of the bound~\eqref{eq:tightbound_Abelian}. As shown in Fig.~\ref{fig:QSLOptimization}, gates whose $SU(2)$ rotation axes are close to the $\sigma_y$ or $\sigma_z$ directions~($\alpha \simeq0,\pi/2$) nearly saturate the Abelian bound. In these regions, the efficiency ratio shows only weak dispersion, indicating that near-optimal solutions are readily found. By contrast, around intermediate orientations,
$\hat{\boldsymbol n}\simeq(\hat{\mathbf y}+\hat{\mathbf z})/\sqrt2$,
the larger spread signals several competing local minima, and the optimization jumps between distinct branches. To elucidate the conditions for maximal efficiency, we analyze representative near-saturating solutions by reconstructing both the candidate contour and the associated optimized surface, and by monitoring the Lie-algebra direction of the transported curvature over $\Sigma(\mathcal C)$. Near saturation, the transported curvature is almost collinear across the optimal surface~(see inset 1), whereas away from saturation it displays pronounced directional variation~(see inset 2). The tighter Abelian bound is therefore approached whenever the curvature becomes effectively Abelian. This indicates that non-Abelianity tends to reduce the QGL-efficiency in the tripod model.

To conclude, we have established quantum geometric limits for
non-Abelian holonomies, bounding the magnitude of arbitrary
Wilczek--Zee holonomies by curvature-norm costs over spanning
surfaces. The associated optimization defines a coupled
contour--surface problem whose boundary dynamics has the form
of a non-Abelian Lorentz force.  In the tripod model, near-optimal
solutions approach QGL saturation by aligning the transported
curvature along a single Lie-algebra direction. These results open
a route toward quantitative geometric-resource bounds for holonomic and topological quantum computation~\cite{zanardi1999,PachosZanardiRasetti2000,Kitaev2003,Nayak2008,Ivanov2001,Alicea2011,DasSarma2015,Zhang2020SOTS,Calzona2020} and for condensed-matter phenomena governed by non-Abelian quantum geometry~\cite{YuQiBernevigFangDai2011,SoluyanovVanderbilt2011,AlexandradinataDaiBernevig2014}.

\noindent \emph{Acknowledgments.}
 We acknowledge support from the Institut Universitaire de France, and from the ANR project QuCoBEC (ANR-22-CE47-0008-02) and from the CAPES-COFECUB (20232475706P) program. F.I. acknowledges support from the Brazilian agencies CNPq (305638/2023-8) and FAPERJ (210.570/2024).

\bibliography{Biblio_QSLNA}

\end{document}